\documentstyle[12pt,epsf]{article} 
\newcommand{\Frac}[2]{\frac{\displaystyle #1}{\displaystyle #2}}
\newcommand{\beq}{\begin{equation}} 
\newcommand{\eeq}{\end{equation}}
\newcommand{\beqn}{\begin{eqnarray}} 
\newcommand{\eeqn}{\end{eqnarray}}
\newcommand{\beqns}{\begin{eqnarray*}}
\newcommand{\eeqns}{\end{eqnarray*}}    
\begin{document}
\begin{titlepage} 
\begin{center}
\hfill INFNNA-IV-2001/12\\ 
\hfill DSFNA-IV-2001/12\\
\hfill hep-ph/0105078\\
\hfill May 2001
\vspace{1.5cm}\\
{\Large\bf The diquark model: New Physics effects for charm and kaon
decays ${}^*$\\}
\vspace*{1.5cm}
{Giancarlo D'Ambrosio$^{\dagger}$ }$\; \; $ and $ \; \; $ {Dao-Neng
Gao$^{\ddagger}$ }
\vspace*{0.4cm} \\ 
{\it Istituto Nazionale di Fisica Nucleare, Sezione di
Napoli, Dipartamento di Scienze Fisiche, Universit\`a di Napoli  
I-80126 Napoli, Italy}
\vspace*{1cm} 
\end{center} 
\begin{abstract}
\noindent
Motivated by diquark exchange, we construct a class of extensions of  
the standard model. These models can generate large CP conserving and CP
violating contributions to the doubly Cabbibo suppressed decays
$D^0\rightarrow K^+\pi^-$ without affecting $D^0-\overline{D^0}$ mixing,
contrary to what is usually believed in the literature. We find an
interesting specific realization of these models, which has the $LR$
chiral structure and can induce novel density $\times$ density
operators. It is new for non-leptonic kaon decays, and particularly, may
provide a possible solution to the $\Delta I=1/2$ rule and direct CP
violation, without inducing large flavour changing neutral currents.  
\end{abstract}

\vfill 
\noindent $^{\dagger}$ E-mail:~giancarlo.dambrosio@na.infn.it\\
\noindent $^{\ddagger}$ E-mail:~gao@na.infn.it, and  
on leave from {\it the Department of
Astronomy and Applied Physics, University of Science and Technology of
China, Hefei, Anhui 230026  China}.\\ \noindent * Work supported in part
by TMR, EC--Contract No. ERBFMRX-CT980169 (EURODA$\Phi$NE). 
\end{titlepage}

\section{Introduction}

$D^{0}-\overline{D^{0}}$ mixing and in general the charm sector is a very
interesting place to test the Standard Model (SM) and its possible
extensions \cite{Buch01, Bigi01, Nirdelta99, Nir99}. Recent data from
FOCUS \cite{focus} and
CLEO \cite{CLEO2000} have given further excitement to this field.  Indeed
the decay $D^{0}\rightarrow K^{+}\pi^{-}$ has been now clearly
observed; this
may occur either through the double Cabibbo suppressed decays (DCS) or $
D^{0}-\overline{D^{0}}$ mixing with a subsequent Cabibbo favoured decays
(CF).\ Data \cite{focus,CLEO2000} seem to exceed the naive SM\
expectation for the ratio of DCS to CF branching fractions is 
\begin{equation}
R_{D}\equiv \left| \frac{A(D^{0}\rightarrow K^{+}\pi ^{-})_{\rm DCS}}{A(%
\overline{D^{0}}\rightarrow K^{+}\pi ^{-})_{\rm CF}}\right| ^{2}
\label{eq:rdcs}
\end{equation}
$\approx $ $\tan ^{4}\theta _{C}\approx $ 0.25\%.\ \ Also SM predictions for 
$D^{0}-\overline{D^{0}}$ mixing give a very negligible contribution.

It is generally believed that extensions of the SM can significantly affect
the mixing but not the decay \cite{Nir99}. We challenge this statement by
constructing a class of models which can generate large CP conserving and
violating contributions only to DCS decays without affecting $D^{0}-
\overline{D^{0}}$ mixing. Thus we want to stimulate the experiments to also
put bounds on CP violating contributions to DCS decays. These models are
obtained by introducing a new scalar particle, a diquark $\chi $, triplet
under colour and can also be theoretically motivated in extensions of the
SM. We fix the $\chi-$coupling so that this is relevant for DCS decays i.e.
\begin{equation}
A(D^{0}\rightarrow K^{+}\pi ^{-})_{\chi }\approx 10^{-2}G_{F}.
\label{eq:DCSchi}
\end{equation}
On the other hand, we also show that a possible large direct CP asymmetry
in $D^\pm$ channel could be induced by the
diquark exchange. 

Further we study more in detail an intriguing specific realization of these
models which may have relevant implications for non-leptonic kaon decays 
\cite{kaon}, in
particular the $\Delta I=1/2$ rule. In fact, diquark interchange may
generate non-leptonic $\Delta I=1/2$ transitions, and the size of these
contributions is appropriately constrained by $\Delta S=2$
interactions. Thus one would have expected that a
sizable interaction like the one in (\ref{eq:DCSchi}) for kaon decays,
though in principle interesting for the $\Delta I=1/2$ rule would generate
a disaster in the flavour changing neutral current (FCNC) sector.\ However
we do find, we think, an elegant solution to this problem, and  
it may also have other applications, for
instance, generating new direct CP violating $\Delta S=1$ operators
without inducing large $\varepsilon$.

Differently from Ref. \cite{glashow} our model is based on supersymmetry 
and we do not address the issue of strong CP problem.

\section{Charm phenomenology}

Mass and width eigenstates of the neutral $D-$ system are written as linear
combinations of the interactions eigenstates:
\begin{equation}
\left| D_{1,2}\right\rangle =p\left| D^{0}\right\rangle \pm q\left| 
\overline{D^{0}}\right\rangle   \label{eq:dmix}
\end{equation}
with eigenvalues $m_{1,2}$ and $\Gamma _{1,2}.\;$The mass and width average
and difference are defined as 
\begin{equation}
m=\frac{m_{1}+m_{2}}{2},\quad \Gamma =\frac{\Gamma _{1}+\Gamma _{2}}{2},
\end{equation}
\begin{equation}
x=\frac{m_{2}-m_{1}}{\Gamma },\quad y=\frac{\Gamma _{2}-\Gamma _{1}}{2\Gamma 
}.
\end{equation}
Decay amplitudes into a final state $f$ are defined by
\begin{equation} 
A_{f}\equiv \langle f\left| H_{W}\right| D^{0}\rangle
,\quad \overline{A_{f}} \equiv \langle f\left| H_{W}\right|
\overline{D^{0}}\rangle .
\end{equation}  
Then one can generally define the complex parameter:
\begin{equation} 
\lambda _{f}\equiv
\frac{q}{p}\frac{\overline{A_{f}}}{A_{f}}.
\end{equation}

The phenomenological evidence of possible large DCS decays comes from the
analysis of CLEO \cite{CLEO2000} and FOCUS \cite{focus} that have been able
to disentangle in the total data sample the DCS contribution from the 
$D^{0}\rightarrow \overline{D^{0}}\rightarrow K^{+}\pi ^{-}$ contribution.
CLEO\ is able to study the time dependence in the ratio of the DCS to the
CF decays, which in the limit of $CP$ conservation is written as:
\begin{eqnarray}
\begin{array}{ll}
R\left( t\right) \equiv  & \left| \frac{\displaystyle \langle K^{+}\pi 
^{-}\left| H_{W}\right| D^{0}(t)\rangle }{\displaystyle \langle K^{+}\pi 
^{-}\left| H_{W}\right| \overline{D^{0}}(t)\rangle }\right| ^{2}=\left[
R_{D}\right.  \\ 
&  \\ 
& \left. +\sqrt{R_{D}}y^{\prime }\Gamma t+\left( y^{\prime 2}+x^{\prime
2}\right) \left( \Gamma t\right) ^{2}/4\right] e^{-\Gamma t},
\end{array}
\label{eq:rws}
\end{eqnarray}
where $R_{D}$ is defined in (\ref{eq:rdcs}) and
the final state interaction will generate a strong phase difference
$\delta$  between the DCS and the CF amplitudes so that the following
rotation is used in (\ref{eq:rws})
\[
\begin{array}{ll}
x^{\prime }= & x\cos \delta +y\sin \delta,  \\ 
y^{\prime }= & y\cos \delta -x\sin \delta. 
\end{array}
\]

CLEO by a careful time dependent study 
finds $R_{D}=$ $\left( 0.33_{-0.065}^{+0.063}\pm 0.040\right) \cdot
10^{-2}$ \cite{CLEO2000} with no-mixing fit, 
while FOCUS \cite{focus} assuming that there is no charm
mixing and no $CP$ violation finds $R_{D}=\left(0.404\pm 0.085\pm
0.025\right) \cdot 10^{-2}$. 
Possible $CP$ violation effects in the mixing, the  direct decay, and 
the interference between those two processes, characterized by $A_M$,
$A_D$, and $\phi$ respectively, can affect the three terms  in
(\ref{eq:rws}),  where, to leading order, both $x^\prime$ and $y^\prime$
are scaled by $(1\pm A_M)^{1/2}$, $R_D\rightarrow R_D(1\pm A_D)$, and
$\delta\rightarrow \delta\pm \phi$, as has been taken into account in the
analysis by CLEO \cite{CLEO2000}.  The corresponding values obtained
in \cite{CLEO2000} are
as follows:
\beq\label{DCPE}
A_M=0.23^{+0.63}_{-0.80},\;\; A_D=-0.01\pm0.17, \;\; 
\sin \phi=0.00\pm0.60.
\eeq

The SM prediction to the $D^0-\overline{D^0}$ mixing is highly suppressed
because it is the second order in $\alpha_W$ and has a very strong GIM
suppression factor $m_s^4/(M_W m_c)^2$. The experimental data are 
\beq\label{expx}
x\leq 0.03 ~\cite{Bigi01},
\eeq
\begin{eqnarray}\label{yprime}
y^\prime \cos\phi=(-2.5^{+1.4}_{-1.6})\cdot 10^{-2}~
\cite{CLEO2000},
\end{eqnarray}
and 
\begin{eqnarray}\label{yfocus} 
y=(3.42\pm1.39\pm0.74)\cdot 10^{-2} ~\cite{focus},
\end{eqnarray}
which cannot be clearly explained in the SM,
where one expects \cite{Buch01}
\beq
x_{\rm SM}, y_{\rm SM}
\le 10^{-3}. 
\eeq

Theoretically, the strong phase $\delta$ was expected small, even
vanishing in the SU(3) limit \cite{Wolf95}. However, as pointed out in
\cite{Nirdelta99}: (i) a large $\delta$ would make the possible different
signs of the measured $D^0-\overline{D^0}$ mixing parameters shown in
eqs. (\ref{yprime}) and (\ref{yfocus}) consistent with each other, and
(ii) recent data allow large values of $\delta$. A large
$\delta$ would be welcome in searching for direct CP asymmetry of
$D^\pm\rightarrow K_S\pi^\pm$.   

At the present, on one hand, the experimental results in charm
phenomenology obviously need to be further improved; on the other hand,
the discrepancy between the SM estimates and the data invites for
speculations about the New Physics contributions. 

\section{Theory}

As mentioned in the Introduction, let us now imagine the theory with the
spin 0 diquark $\chi$ with the quantum numbers as follows
\cite{Zwirner, ellis, glashow},
i.e. 
\[
\begin{array}{ccccc}
& SU(3)_C & SU(2)_L & U(1)_Y & B\\ 
\chi & \mathbf{3} & 1 & -1/3 &-2/3\\ 
\chi^{c} & \overline{\mathbf{3}} & 1 & 1/3& 2/3
\end{array}
\]
coupled to quark left-handed doublets, $Q$ and right--handed singlets $
U,D$ and assume now that these are supersymmetric degrees of freedom. We
write the R-parity conserving interaction in the superpotential $W$
\cite{Zwirner, ellis}
\begin{equation}
W_{diquark}=g_{L}\left( h_{L}\right)^A_{ij}Q^{i}Q^{j}\chi _{A}+g_{R}\left(
h_{R}\right)^A_{ij}U_{i}^{c}D_{j}^{c}\chi _{A}^{c},  \label{Wsuper}
\end{equation}
where $i,j$, and $A$ are family indices, and the possible
intergenerational mixing in $\left(h_{L}\right)^A_{ij}$ and
$\left(h_{R}\right)^A_{ij}$ is
assumed. $h^A_{L}$ are flavour symmetric matrices in the
weak-isospin basis, but they depart from symmetry in mass-eigenstate
basis.  Also (super-)Yukawa couplings of quarks to the two Higgses belong
to this so-called superpotential.\ All this does make sense in the
supersymmetric version of $E_{6}$ \cite{HR89},  therefore the terms 
in the superpotential are protected by the no-renormalization
theorem \cite{martin99}.  Anyway, the supersymmetric predictions, for our
purpose, can be regarded as the effective predictions of extensions of the
SM satisfying the phenomenological limits including LEP data. This is
different from Ref. \cite{glashow} which is in the
framework of a standard renormalizable theory.

From eq. (\ref{Wsuper}), one can get the four-quark operators
which have $LL$ and $RR$ chiral structures mediated by the diquarks $\chi$
and $\chi^c$ respectively. Interestingly, $\chi-\chi^c$ mixing
\cite{Zwirner} will generate instead the $LR$ chiral structure.  
Generally, squaring $\Delta S=1$ operators \cite{kaon}  could
automatically lead to
dangerous FCNC transitions.  However, this is not the
case for $\Delta C=1$ operators contributing to DCS $D\rightarrow
K\pi$ decays. Therefore, as shown in the next section, 
all the chiral structures $LL$, $RR$, and $LR$ can  
enhance the DCS decays without affecting $D^0-\overline{D^0}$
mixing, while only using the $LR$ structure, one can get the possible
large contributions to $\Delta I=1/2$ transitions and new direct CP
violation without large FCNC in kaon sector. 

As further motivation, Voloshin recently \cite{vol00} has considered a
new centiweak four-quark interaction, with the strength $10^{-2} G_F$ to
reproduce the experimental ratio of $\tau(\Lambda_b)/\tau(B_d)$. The new
interaction arises through a weak $SU(2)$ singlet scalar field with
quantum numbers of diquark $\chi$:
\beq 
b_R u_R\rightarrow\chi\rightarrow c_L d_L,
\eeq
and the chiral structure is like
the $LR$ one of this paper.

\section{Phenomenological analysis}

\subsection{ DCS $D\rightarrow K\pi$ decays and $D^0-\overline{D^0}$
mixing}

It is straightforward to get the following chiral structures which could
contribute to the DCS decays $D\rightarrow K\pi$:
\begin{equation}\label{LL}
{\cal L}^{LL}=\frac{g_L^2}{2
m_\chi^2}(h_{11}^{L*}h_{22}^L)\left[(\overline{u_L}\gamma_\mu
c_L)(\overline{d_L}\gamma^\mu s_L)-(\overline{u_L}\gamma_\mu
s_L)(\overline{d_L}\gamma^\mu c_L)\right]+h.c., 
\end{equation}
\begin{equation}\label{RR}
{\cal L}^{RR}=
\frac{g_R^2}{2m_\chi^2}(h_{11}^{R*}h_{22}^R)\left[(\overline{u_R}\gamma_\mu
 c_R)(\overline{d_R}\gamma^\mu s_R)-(\overline{u_R}\gamma_\mu
s_R)(\overline{d_R}\gamma^\mu c_R)\right]+h.c.,
\end{equation}
and
\begin{eqnarray} \label{LR}
{\cal L}^{LR}&=&\frac{g_Lg_R}{2
m_\chi^2}(h_{11}^{R*}h_{22}^L)\left\{\left[(\overline{u_R}c_L)
(\overline{d_R}s_L)-(\overline{u_R}s_L)(\overline{d_R}c_L)\right]
\right. \nonumber \\
&&\left. +\frac{1}{4}\left[(\overline{u_{R}}\sigma ^{\mu \nu
}c_{L})(\overline{d_{R}}\sigma _{\mu \nu }s_{L})-(\overline{u_{R}}\sigma
^{\mu \nu }s_{L})(\overline{d_{R}}\sigma_{\mu
\nu}c_{L})\right]\right\}+h.c.
\end{eqnarray}
Note that the $LR$ structure (\ref{LR}) is derived by assuming the mixing
between $\chi$ and $\chi^c$, and we neglect the tensor contributions in
the present work. 
Typically we can choose
\begin{equation}\label{input}
\frac{g_L^2}{m_\chi^2}=\frac{g_R^2}{m_\chi^2}=\frac{g_L g_R}{m_\chi^2}\sim
10^{-6}~{\rm GeV^{-2}} 
\end{equation}
for $m_\chi=300~{\rm GeV}$,  and $g_L=g_R=0.3$.
Thus, we can fix $h_{11}^{L*}h_{22}^L$, $h_{11}^{R*}h_{22}^R$, and
$h_{11}^{R*}h_{22}^L$ to render that the diquark
contributions to the DCS decays $D^0\rightarrow K^+\pi^-$ from
eqs. (\ref{LL}), (\ref{RR}), and (\ref{LR}) separately satisfy 
\begin{equation}
A(D^0\rightarrow K^+\pi^-)_\chi=G_\chi \approx 10^{-2} G_F,
\end{equation}
which can compete with the corresponding SM contribution.  

Note that all the couplings in eqs. (\ref{LL}), (\ref{RR}), and
(\ref{LR}) do not induce $\Delta C=2$ transitions, therefore, in the
diquark models, one can get the enhancement of DCS decays $D^0\rightarrow
K^+\pi^-$ without the large $D^0-\overline{D^0}$ mixing.
Since $D^0-\overline{D^0}$ mixing will involve 
other matrix elements of $h_L$ and $h_R$ than $h_{11}^{L, R}$ and
$h_{22}^{L, R}$, we can tune these new couplings to accommodate the
experimental bounds of this mixing.

The new $\Delta C=1$ dynamics induced from the structures in
eqs. (\ref{LL})-(\ref{LR}) can contribute to the direct CP violation in
DCS $D^0\rightarrow K^+\pi^-$, which is
\beq
A_D^\chi\sim \Im m (h_{11}^{A*}h_{22}^B)\sin\delta,
\eeq
where $A=B=L$ denotes the contribution from eq. (\ref{LL}), $A=B=R$ from
eq. (\ref{RR}), and $A=R$ and $B=L$ from
eq. (\ref{LR}) respectively.  From
experimental value in eq. (\ref{DCPE}),  $A_D=-0.01\pm 0.17$, we can get 
\beq\label{phase} 
|\Im m (h_{11}^{A*}h_{22}^B)\sin\delta|<0.2.
\eeq

The charge asymmetry in $D^\pm\rightarrow K_S \pi^\pm$  arises
from the interference between the CF $D^\pm\rightarrow \overline{K^0}\pi^\pm$ 
and DCS $D^\pm\rightarrow K^0\pi^\pm$ decays, and
the $K^0-\overline{K^0}$ mixing will give the following contribution
without any theoretical uncertainty \cite{Bigi01,BY95}:
\beqn\label{DCP}
\Frac{\Gamma(D^+\rightarrow K_S\pi^+)-\Gamma(D^-\rightarrow K_S \pi^-)}{
\Gamma(D^+\rightarrow K_S\pi^+)+\Gamma(D^-\rightarrow K_S \pi^-) }=-2 Re
~\varepsilon_K \simeq -3.3\cdot 10^{-3}.
\eeqn
Here the same asymmetry both in magnitude and in sign as
eq.(\ref{DCP}) will arise for the final state with a $K_L$ instead of
a $K_S$.
On the other hand, from
eqs. (\ref{LL})--(\ref{LR}), one can get contribution to the asymmetry
from the diquark exchange as
\beq
\Frac{\delta \Gamma_\chi}{2\Gamma}=\Frac{|\Gamma(D^+\rightarrow
K_S\pi^+)-\Gamma(D^-\rightarrow K_S \pi^-)|_\chi}{\Gamma(D^+\rightarrow
K_S\pi^+)+\Gamma(D^-\rightarrow K_S \pi^-) }
\sim |\Im m
(h_{11}^{A*}h_{22}^{B})\sin\delta |\cdot 10^{-1}.
\eeq
The factor $10^{-1}$ is due to the ratio of $\Frac{g_A g_B}{m_\chi^2}$ and
$G_F$. Thus using eq. (\ref{phase}), one can get 
\beq
\Frac{\delta \Gamma_\chi}{2\Gamma}\le 10^{-2}.
\eeq
Here eqs. (\ref{LL})--(\ref{LR}) will make a contribution of the opposite
sign to the asymmetry in $D^+\rightarrow K_L\pi^+$ vs. $D^-\rightarrow
K_L\pi^-$, which is different from the case of eq. (\ref{DCP}).
Note that this upper bound is one order larger than the value given in
(\ref{DCP}), which is consistent with the statement in Ref. \cite{BY95}. 
Therefore, it is of interest to carry out the precise
measurement of this asymmetry in order to exploit New Physics
effects.

It is found that all three chiral structures $LL$, $RR$, and $LR$
[(\ref{LL})--(\ref{LR})] can separately produce the large contributions to
DCS $D^0\rightarrow K^+\pi^-$ decays without affecting  
$D^0-\overline{D^0}$ mixing. This is somewhat contrary to the statement in
Ref. \cite{Nir99}. In the kaon physics, only $LR$ chiral structure is
useful when we consider the constraints from the $K^0-\overline{K^0}$
mixing.
 
\subsection{$K^0-\overline{K^0}$ mixing, $\Delta
I=1/2$ rule, and direct $CP$ violation} 

Diquark exchange between the  LR structure generates 
\begin{eqnarray}
{\cal L}^{LR}=\frac{g_L g_R}{2 m_{\chi }^{2}}\left\{
(h_{11}^{R*}h_{12}^L)\left[(\overline{u_R} u_L)(\overline{d_R}s_L)
-(\overline{u_{R}}s_{L})(\overline{d_{R}}u_{L})\right]\right.\nonumber\\
+\left.(h_{12}^{R}h_{11}^{L*})\left[(\overline{u_L}u_R)(\overline{d_L}s_R)-
(\overline{u_{L}}s_{R})(\overline{d_{L}}u_{R})\right] \right\}+h.c. 
\label{1/2}
\end{eqnarray}
The matrix elements of these operators can be enhanced compared to the
usual $Q_{-} $ operator \cite{kaon}, and they will induce pure
$\Delta I=1/2$ transitions. Therefore, if $h_{ij}^{L(R)}$'s appearing in
eq. (\ref{1/2}) are not very small, one can expect a possible solution to
the $\Delta I=1/2$ rule. 

However, we have to  show that this structure could avoid large
FCNC. Indeed $K^0-\overline{K^0}$ mixing can be generated from $\left(
\overline{d_{R}}s_{L}\right)^{2}$, $\overline{
d_{R}}s_{L}$ $\overline{d_{L}}s_{R}$, and $\overline{d_{L}}\gamma ^{\mu
}s_{L}$ $\overline{d_{L}}\gamma_{\mu}s_{L}.\;$
The last one is generated by the
usual $Q_{-}$ operator. If we assume that only one of the
terms in (\ref{1/2}),
for instance $h_{11}^{R*}h_{12}^{L},$ is large and the other is very
small then squaring the structure in (\ref{1/2}) will \textbf{not} 
generate $K^0-\overline{K^0}$ mixing because  
\[
\left\langle \overline{u_{R}}u_{L}\overline{u_{R}}u_{L}\right\rangle =0. 
\]

Also, if we assume some electroweak phases in the  
${\Delta S=1}$ transitions induced by the diquark, we
can obtain the contribution to $\varepsilon _{\chi }^{\prime }$. The only
thing we have to be concerned that we do not generate too much $\varepsilon
_{\chi },$ i.e. $H_{\Delta S=2,CP}^{\chi }$ larger than the one in SM. 
Indeed in the SM 
\[
\Re e(\varepsilon )\sim \frac{\Im m\left\langle \overline{K}|H_{\Delta
S=2}|K\right\rangle }{\Re e\left\langle \overline{K}|H_{\Delta
S=2}|K\right\rangle }\sim 2\cdot10^{-3}. 
\]
The diquark exchange can generate $\Im m(A_{0})$ in 
\[
\varepsilon^{\prime }=i{\frac{e^{i(\delta _{2}-\delta
_{0})}}{\sqrt{2}}}\omega
\left[ {\frac{\Im m(A_{2})}{\Re e(A_{2})}}-{\frac{\Im m(A_{0})}{\Re
e(A_{0})\,}}\right], 
\]
to match the experimental result
$\Frac{\varepsilon^{\prime}}{\omega}\sim 10^{-4}$ \cite{Bigi01, kaon} with
a value for the imaginary part \begin{equation}
\Im m(h_{12}^{L}h_{11}^{R*})\sim 10^{-3}.  \label{CPchi}
\end{equation}
Now since we claim that with a particular choice of $h^{R}$ and $h^{L}$,
\[
\Re e\left\langle \overline{K}|H_{\Delta S=2}^{\chi }|K\right\rangle <\Re
e\left\langle \overline{K}|H_{\Delta S=2}^{SM}|K\right\rangle 
\]
then with the value in (\ref{CPchi}) we obtain 
\[
\frac{\Im m\left\langle \overline{K}|H_{\Delta S=2}|K\right\rangle }{\Re
e\left\langle \overline{K}|H_{\Delta S=2}|K\right\rangle }<10^{-3} 
\]
and so there is no problem for $\Re e(\varepsilon).$  If the electroweak
phase is only in $h^L_{12}$, the induced electric dipole moment of the
neutron is smaller than the experimental value \cite{RGF00}.

\subsection{The diquark is coupled to the first two
 generations}

In this subsection, we present an example to show that one can address 
simultaneously the issue of DCS $D^0\rightarrow K^+\pi^-$ decays,
the contributions to $\Delta I=1/2$ transitions and the direct CP
violation in kaon sector without large FCNC. 
For simplicity, we assume 
the following 2$\times$2 matrices for $h^R$ and $h^L$ 
\begin{equation}
h^{R}=\left( 
\begin{array}{cc}
1 & \lambda ^{4}   \\ 
\lambda ^{2} & \lambda ^{2}   
\end{array}
\right) ,\quad h^{L}=\lambda\left( 
\begin{array}{ccc}
1 & 1   \\ 
1 & 1   
\end{array}
\right),
\label{textures}
\end{equation}
i.e. the diquark is coupled to the first two generation quark fields.
Note that the matrix elements of $h_L$ and $h_R$ merely have the
meaning of order of magnitude, 
therefore, it should be understood, for instance, $h^L_{11}\sim
O(\lambda)$, and $\lambda=0.22$ is the Wolfenstein parameter. So our
analysis is at qualitative level.

From (\ref{textures})
$h_{11}^{L*}h_{22}^L\sim h_{11}^{R*}h_{22}^R\sim O(\lambda^2)$, and
$h_{11}^{R*}h_{22}^L\sim O(\lambda)$, so induced by the diquark, one can
obtain the DCS $D^0\rightarrow K^+\pi^-$ decays amplitude which can be
compared with the SM contribution, as shown in eq. (\ref{eq:DCSchi}). 

\begin{figure}
\begin{center}
\leavevmode
\hbox{
\epsfxsize=8cm
\epsffile{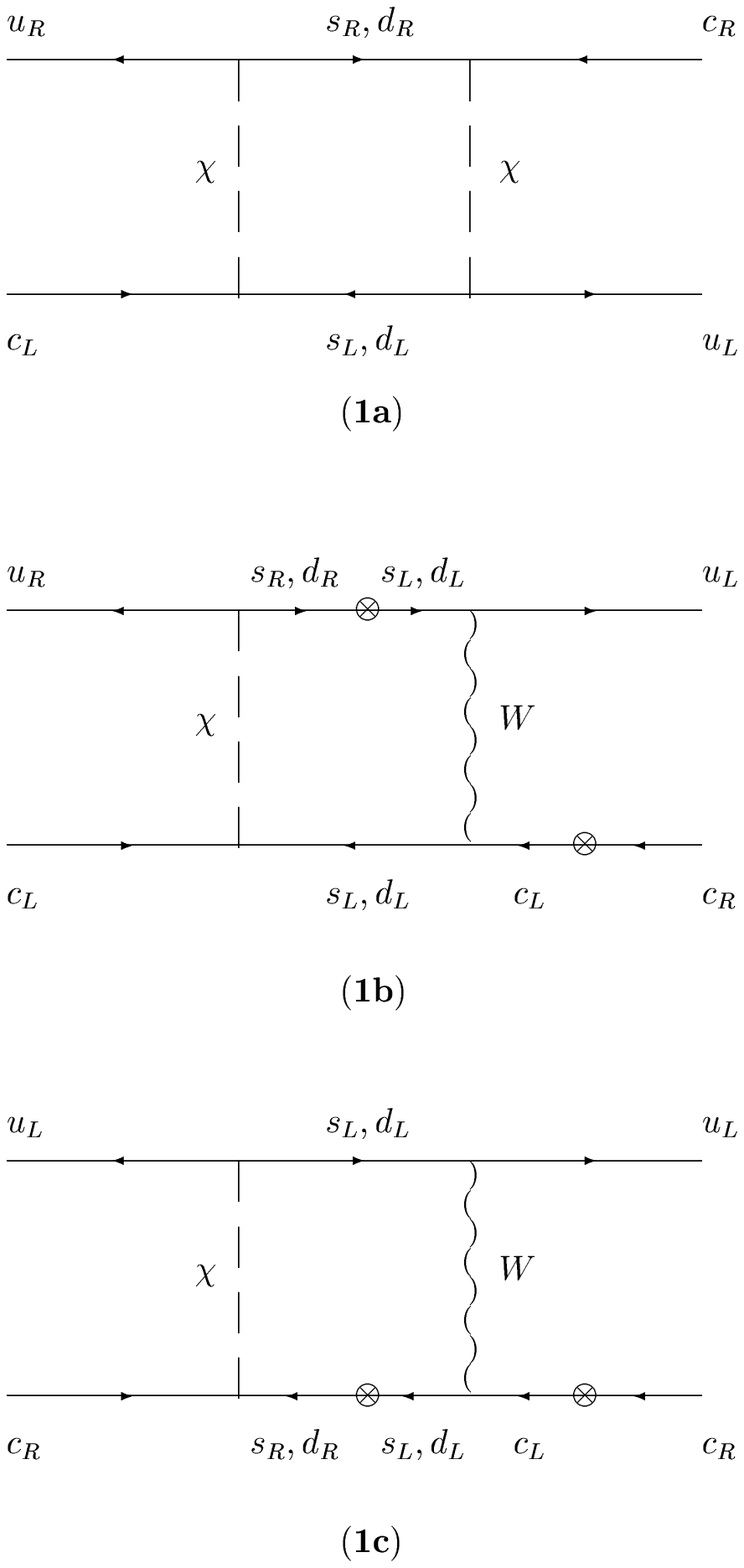}}
\end{center}                       
\caption{The box diagrams contributing to $D^0-\overline{D^0}$
mixing. The $\otimes$ denotes the chirality flip on the internal
and external lines. 
} 
\end{figure} 

Now we check the $D^0-\overline{D^0}$ mixing generated by the diquark
exchange. The most relevant box diagrams are $\chi-\chi-$box and $\chi-W-$
box, which have been drawn in Figure 1. Here, we only
consider the contributions induced by the $LR$ chiral
structure, and the calculation is straightforward. From Fig. (1a), one can
get
\beq
\xi
(h_{22}^{L*}h_{12}^{L}+h_{21}^{L*}h_{11}^{L})\cdot(h_{21}^{R*}h_{11}^{R}+   
h_{12}^{R*}h_{22}^R)~(\overline{u_{R}}c_{L})(\overline{u_{L}}c_{R}),
\eeq
where
\begin{equation} 
\xi=\frac{g_{L}^{2}g_{R}^{2}}{16\pi ^{2}m_{\chi 
}^{2}}\sim 10^{-10}~{\rm GeV}^{-2}
\end{equation}   
for the same values of $g_L$, $g_R$, and $m_\chi$ used in
eq. (\ref{input}). Likewise, the contributions of Fig. (1b) and
Fig. (1c) are respectively
\beqn
\frac{G_{F}}{\sqrt{2}}\frac{g_{L}g_{R}m_{c}}{4\pi^{2}m_{\chi }^{2}} 
(V_{cd}h_{21}^{L}+V_{cs}h_{22}^{L})\cdot 
(m_{d}h_{11}^{R*}V^*_{ud}+m_{s}h_{12}^{R*}V^*_{us})
~(\overline{u_{R}}c_{L})(\overline{u_{L}}c_{R}), 
\eeqn
and
\beqn
\frac{G_{F}}{\sqrt{2}}\frac{g_{L}g_{R}m_{c}}{4\pi
^{2}m_{\chi }^{2}} (V^*_{ud}h_{11}^{L*}+V^*_{us}h_{12}^{L*})\cdot 
(m_{d}h_{21}^{R}V_{cd}+m_{s}h_{22}^{R}V_{cs}) 
~(\overline{u_{L}}c_{R})(\overline{u_{L}}c_{R}).
\eeqn

Experimentally, as shown in eq. (\ref{expx}), $x\leq 0.03$, and the SM
predicts $x_{\rm SM}\sim 10^{-3}$.
Using eq. (\ref{textures}), we can get that all the above box
diagrams lead to $x_\chi\sim 10^{-3}-10^{-2}$, not larger than the
experimental value.

\begin{figure} 
\begin{center}
\leavevmode
\hbox{
\epsfxsize=8cm
\epsffile{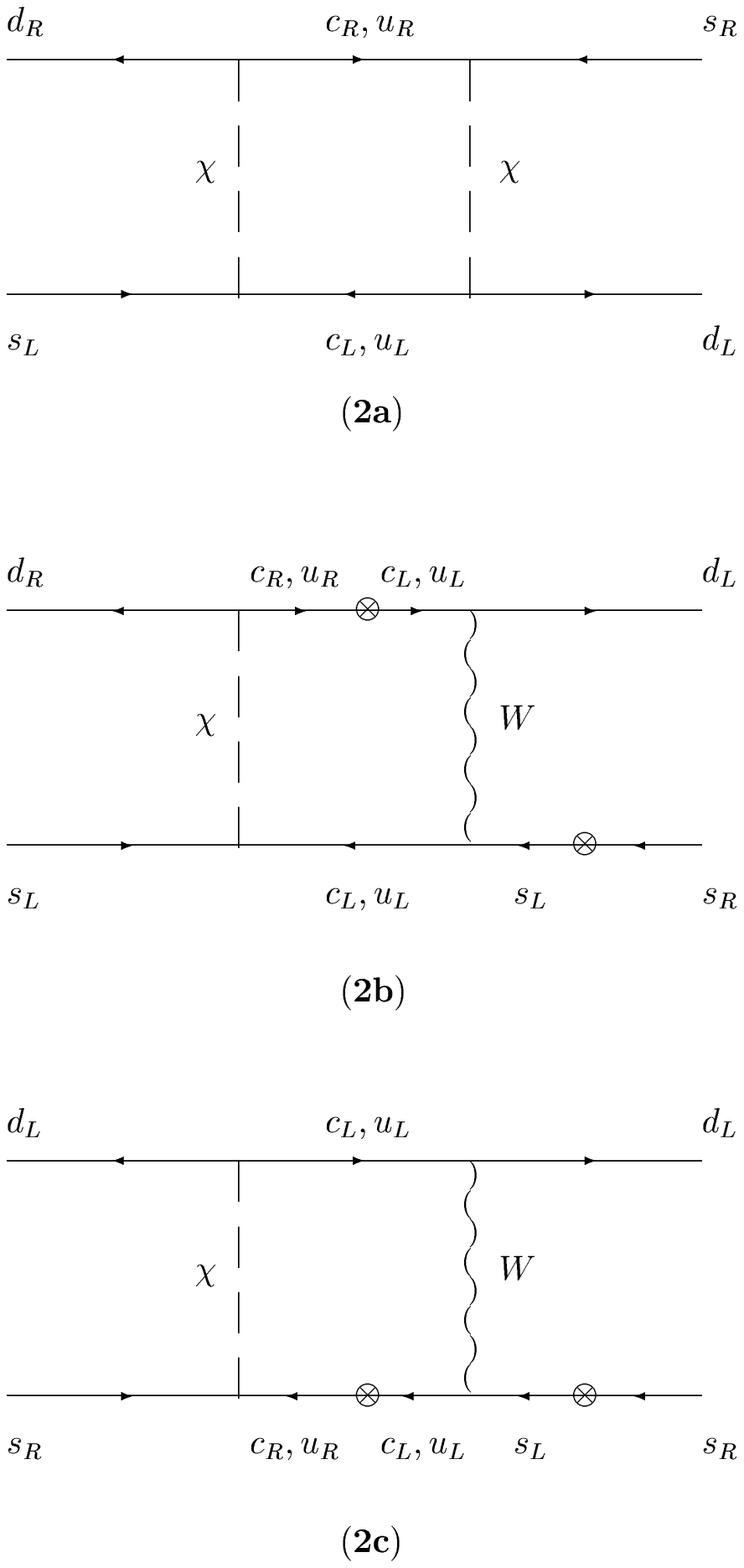}}
\end{center}  
\caption{The box diagrams contributing to $K^0-\overline{K^0}$
mixing. The $\otimes$ denotes the chirality flip on the internal and
external lines.} 
\end{figure} 

From eq. (\ref{textures}) $h^{R*}_{11}h^L_{12}\sim O(\lambda)$ and thus
the amplitudes of non-leptonic $\Delta I=1/2$ transitions of kaon decays
could be $10^{-2} G_F$.
In order to check the problem of FCNC, similar box
diagrams, which are shown in Figure 2,  have been calculated. 
Fig. (2a), Fig. (2b) and Fig. (2c) will give  respectively the following
contributions 
\beq
\xi
(h_{11}^{L*}h_{12}^{L}+h_{21}^{L*}h_{22}^{L})
\cdot(h_{12}^{R*}h_{11}^{R}+h_{12}^{R*}h_{22}^{R}) 
~(\overline{d_{L}}s_{R})(\overline{d_{R}}s_{L}),
\eeq
\beq
\frac{G_{F}}{\sqrt{2}}\frac{g_{L}g_{R}m_{s}}{4\pi
^{2}m_{\chi }^{2}}
(V_{us}h_{12}^{L}+V_{cs}h_{22}^{L})\cdot  
(m_{c}h_{21}^{R*}V^*_{cd}+m_{u}h_{11}^{R*}V^*_{ud}) 
~(\overline{d_{L}}s_{R})(\overline{d_{R}}s_{L}), 
\eeq
and
\beq
\frac{G_{F}}{\sqrt{2}}\frac{g_{L}g_{R}m_{s}}{4\pi
^{2}m_{\chi }^{2}}
(V^*_{ud}h_{11}^{L*}+V^*_{cd}h_{21}^{L*})\cdot       
(m_{u}h_{12}^{R}V_{us}+m_{c}h_{22}^{R}V_{cs}) 
~(\overline{d_{L}}s_{R})(\overline{d_{L}}s_{R}). 
\eeq

It is easy to find that all the above contributions are not larger than
the SM prediction of the $K^0-\overline{K^0}$ mixing. 

Another FCNC problem could be $Z^0$ penguin diagram contribution
to the decay $K_L\rightarrow\mu\overline{\mu}$. It will generate the
following effective hamiltonian 
\beq
\Frac{g_Lg_R\alpha_{\rm EM}m_c m_\mu}{4\pi m_\chi^2 m_W^2 {\rm
sin}^2\theta_W}(h_{22}^{L*}h_{21}^R+h_{21}^L
h_{22}^{R*})\overline{d}\gamma_5 s\overline{\mu}\gamma_5\mu,
\eeq
which has to be  compared with the SM prediction
$\sim \lambda^2 (10^{-12} {\rm GeV}^{-2}) \overline{d}\gamma_5 s
\overline{\mu}\gamma_5\mu$
and thus substantially smaller.

Furthermore,  if we put a small electroweak phase $\varphi\sim 10^{-2}$ in
$h^L_{12}$, $\Im m (h^L_{12}h^{R*}_{11})\sim \lambda {\rm sin}\varphi\sim
10^{-3}$ and eq. (\ref{CPchi}) will hold.  
Also as pointed out in the previous subsection, 
no large electric dipole moment of the neutron will be induced.

It has been shown that, in the simple realization (\ref{textures}) with
only one diquark, we get the enhancement of the amplitudes of DCS
$D^0\rightarrow K^+\pi^-$ decays and non-leptonic $\Delta I=1/2$
transitions of kaon decays up to $10^{-2} G_F$ without any dangerous FCNC.     
Phenomenologically we could also make $h_{12}^L\sim O(1)$ and not
$O(\lambda)$ as in eq. (\ref{textures}), which means that a larger
contribution to $\Delta I=1/2$ kaon decays would be possible. However, we
do not want $h^L$ in (\ref{textures}) to depart severely from a
symmetric structure, but a larger enhancement could be still achieved from
the hadronic matrix element.

\section{Conclusions}

In this paper, we have constructed a class of models motivated by diquark
exchange, which can generate large contributions to the DCS
$D^0\rightarrow K^+\pi^-$
decays without affecting $D^0-\overline{D^0}$ mixing. Our conclusion 
somewhat disagrees with the statement 
in Ref. \cite{Nir99} that the New Physics can only affect
significantly the mixing but not the decay. A large direct CP
asymmetry in $D^\pm\rightarrow K_S\pi^\pm$ is possible in our model, which
may be regarded as the signal to look for New Physics scenarios.

All the chiral structures including $LL$, $RR$, and $LR$ can lead to the
enhancement of the DCS decays in the charm sector, however, only $LR$
structure is useful in kaon sector when we impose the constraints by
avoiding the large FCNC. It is particularly interesting that
this $LR$ structure can generate novel density $\times$ density operators,
which can induce
the pure $\Delta I=1/2$ transitions and new direct CP violation. 
To our knowledge, the role of these operators in non-leptonic kaon decays
is discussed for the first time in the present paper.  

\vspace{1.2cm}
\begin{center}
{\bf ACKNOWLEDGMENTS}
\end{center}
G.D. wishes to thank Shelly Glashow for drawing his attention on 
diquark contributions in weak interactions, Riccardo Barbieri
and Gino Isidori for very illuminating
discussions, and the support of the ``Bruno Rossi" INFN-MIT exchange
program.

\end{document}